\documentstyle{l-aa}

\newcommand{\apj}{{ApJ\ }}
\newcommand{\aap}{{A\&A\ }}
\newcommand{\mnras}{{MNRAS\ }}
\newcommand{\aj}{{AJ\ }}
\newcommand{\apjs}{{ApJS\ }}
\newcommand{\apjl}{{ApJ Lett.\ }}

\begin{document}

\thesaurus{03(11.02.2; 13.07.2)}

\title{TeV gamma-ray observations of Southern BL Lacs with the\\
    CANGAROO 3.8m Imaging Telescope}

\author{M.D. Roberts\inst{1,2}, S.A. Dazeley\inst{2}, 
P.G. Edwards\inst{3}, T. Hara\inst{4}, Y. Hayami\inst{5}, 
J. Holder\inst{1}, F. Kakimoto\inst{5}, S. Kamei\inst{5}, 
A. Kawachi\inst{1}, T. Kifune\inst{1}, R. Kita\inst{6}, 
T. Konishi\inst{7}, A. Masaike\inst{8}, Y. Matsubara\inst{9}, 
T. Matsuoka\inst{9}, Y. Mizumoto\inst{10}, M. Mori\inst{1}, 
H. Muraishi\inst{6}, Y. Muraki\inst{9}, K. Nishijima\inst{12}, 
S. Oda\inst{7}, S. Ogio\inst{5}, J.R. Patterson\inst{2}, 
G.P. Rowell\inst{1,2}, T. Sako\inst{9}, K. Sakurazawa\inst{5}, 
R. Susukita\inst{13}, A. Suzuki\inst{7}, R. Suzuki\inst{5}, 
T. Tamura\inst{14}, T. Tanimori\inst{5}, G.J. Thornton\inst{2}, 
S. Yanagita\inst{6}, T. Yoshida\inst{6} and T. Yoshikoshi\inst{1}}

\offprints{M.D. Roberts}

\institute{Institute for Cosmic Ray Research, University of Tokyo,
  Tokyo 188, Japan
 \and Department of Physics and Mathematical Physics, University of
  Adelaide, South Australia 5005, Australia
 \and Institute of Space and Astronautical Science, Kanagawa 229,
  Japan
 \and Faculty of Commercial Science, Yamanashi Gakuin University,
  Yamanashi 400, Japan
 \and Department of Physics, Tokyo Institute of Technology,
  Tokyo 152, Japan
 \and Faculty of Science, Ibaraki University, Ibaraki 310, Japan
 \and Department of Physics, Kobe University, Hyogo 637, Japan
 \and Department of Physics, Kyoto University, Kyoto 606, Japan
 \and Solar-Terrestrial Environment Laboratory, Nagoya University,
  Aichi 464, Japan
 \and National Astronomical Observatory of Japan, Tokyo 181, Japan
 \and Faculty of Education, Miyagi University of Education,
  Miyagi 980, Japan
 \and Department of Physics, Tokai University, Kanagawa 259,
  Japan
 \and Institute of Physical and Chemical Research, Saitama 351-01,
  Japan
 \and Faculty of Engineering, Kanagawa University, Kanagawa 221,
  Japan}

\maketitle
\markboth{Roberts et al.}{TeV Observations of Southern BL~Lacs}
%\maintitlerunninghead{TeV Observations of Southern AGN}
%\authorrunninghead{Roberts et al.}
\begin{abstract}

Observational and theoretical results indicate that low-redshift
BL~Lacertae objects are the most likely extragalactic sources
to be detectable at TeV energies.
In this paper we present the results of 
observations of 4 BL~Lacertae objects 
(PKS0521$-$365, EXO\,0423.4$-$0840, PKS2005$-$489 and PKS2316$-$423)
made between 1993 and 1996 with the
CANGAROO 3.8\,m imaging \v Cerenkov telescope. During the period
of these observations the gamma-ray energy
threshold of the 3.8m\,telescope was $\sim$2\,TeV.  Searches for
steady long-term emission have been made, and, inspired by the
TeV flares detected from Mkn~421 and Mkn~501, a search on a
night-by-night timescale  has also been performed for each source. 
Comprehensive Monte Carlo simulations are
used to estimate upper limits for both steady and short timescale
emission.


\end{abstract}

\section{Introduction}

Only two extragalactic sources have been confirmed to emit 
gamma-rays
above 300\,GeV: Mkn~421 (\cite{pun92}, \cite{mac95}, \cite{pet96}) 
and Mkn~501 (\cite{qui96}, \cite{aha97}).
In addition, an as yet unconfirmed detection has been made of
1ES~2344+514 (\cite{cat97a}). All three have redshifts less than 0.05
and are, based on their values of log~($F_X$/$F_r$), classified
as X-ray--selected BL~Lacs (XBLs). Observations of more than 30 
candidates of other source types over a range of redshifts have not yielded
any other extragalactic TeV emitters (e.g. \cite{ker95}).
%The AGN detected at energies above 100\,MeV by EGRET (e.g. Mattox
%et al. 1997 and references therein) were initially thought to
%be good candidates for TeV observations.
From theoretical considerations Stecker et al. (1996) have proposed
that low redshift XBLs may be the only extragalactic gamma-ray sources
observable at TeV energies. 

The CANGAROO data set, containing data collected over more than 5 years, 
includes observations of four
low redshift active galactic nuclei (AGN) of 
the BL~Lacertae class (PKS0521$-$365, EXO\,0423.4$-$0840,
 PKS2005$-$489 and PKS2316$-$423). In this paper we present the results
of searches for TeV emission from these objects.  
A comparison of on-source and off-source regions of sky is used
to search for an excess of gamma-ray--like
events from these sources over a typically two-week observing period.

The detection of a large flare from the BL~Lac Mkn~421 by the Whipple
Collaboration (\cite{gai96}) has encouraged us to
analyze the data
with particular emphasis on shorter timescale flare searches. The Whipple 
result has shown that at TeV energies, BL~Lacs are capable of 
extremely energetic flares on timescales of less than 1~day. 
Preliminary analysis of the Mkn~421 gamma-ray spectrum indicates that
photon energies extend beyond 5\,TeV (\cite{mce97}). Similar strong,
short duration flares at TeV energies have also been reported from 
Mkn~501, showing gamma-ray
emission extending to energies of at least 10\,TeV (\cite{aha97,qui97}).
The lack of high energy turnover in the
observed spectrum implies that the 
interaction of the gamma-rays with the
cosmic infra-red background is at the lower end of expectations 
(cf \cite{ste92}). The results from Mkn~421 and Mkn~501
indicate that the CANGAROO 3.8\,m telescope, with a relatively high
gamma-ray energy threshold, is capable of detecting such extragalactic
sources of TeV gamma-rays.

\section{The CANGAROO imaging atmospheric \v Cerenkov telescope}

The CANGAROO 3.8\,m imaging telescope is located near Woomera, South Australia
(longitude $137^{\circ}47'$E, latitude $31^{\circ}06'$S, 160\,m a.s.l).
The reflector is a single 3.8\,m diameter parabolic dish with $f$/1.0 optics.
The imaging camera consists of a square-packed array of 10mm $\times$ 10mm
Hamamatsu R2248 photomultiplier tubes. The camera originally contained
224 photomultiplier tubes, and was increased to 256 pixels 
(16 $\times$ 16 square
array) in May 1995. The tube centers are separated by $0.18^{\circ}$, giving
a total field of view (side-side) of $\sim 3.0^{\circ}$. The photo-cathode of
each tube subtends $0.12^{\circ} \times 0.12^{\circ}$, giving a
photo-cathode coverage of about 40\%
of the field of view. 
For a more detailed description of the 3.8\,m
telescope see \cite{har93}.

In the current configuration an event trigger is generated when a 
sufficiently large number of tubes (3$\sim$5) exceed their 
discriminator threshold.
Individual tube discriminator levels are believed to be around
4 photo-electrons. Under these triggering conditions the current gamma-ray
energy threshold of the 3.8\,m telescope is estimated to be $\sim$1.5\,TeV. 
Prior
to mirror re-coating in November 1996 (which includes all data presented
in this paper) the energy threshold was somewhat higher. Using Monte Carlo
simulations we estimate that this energy threshold (as defined by the 
peak of the differential energy spectrum) was 2.5\,TeV. When 
calculating integral fluxes and flux limits we define our threshold as 2\,TeV.
This figure is obtained by re-binning the lower energy gamma-rays to 
produce an integral spectrum defined by a single power law with a sharp
cutoff at the threshold energy.

Since starting observations in 1991 the CANGAROO 3.8\,m telescope has been
used to observe a number of galactic and extragalactic candidate TeV
gamma-ray sources (\cite{rob97}). From these observations we have
evidence for gamma-ray emission from three galactic sources --- PSR1706$-$44
(\cite{kif95}), the Crab nebula (\cite{tan94}) and
the Vela SNR (\cite{yos97}).

As an indication of the sensitivity of the 3.8\,m telescope
to extragalactic sources we can use Monte Carlo simulations to 
estimate the response of the telescope to fluxes observed from
Mkn~421 and Mkn~501.
Assuming that the gamma-ray emission from Mkn~421 and Mkn~501 extends
up to 10\,TeV  
the integral fluxes above 2\,TeV are as follows.

For Mkn~421:\\
F($>$2{\rm TeV}) $\sim 2.2 \times 10^{-12}$ photons ${\rm cm}^{-2}{\rm s}^{-1}$\\
(adopting the average 1996 Whipple flux with an assumed integral spectral index
of $-$1.56, \cite{mce97}.)

For Mkn~501:\\
 F($>$2{\rm TeV}) $\sim 6.5 \times 10^{-12}$ photons ${\rm cm}^{-2}{\rm s}^{-1}$\\
(adopting the March 1997 HEGRA flux assuming an integral photon spectral
index of $-$1.49, Aharonian et al. 1997.) 

If any of the sources examined in this paper were 
capable of providing a steady 
%DC 
flux at the level of Mkn~421 or Mkn~501 it would
be detectable by the CANGAROO 3.8\,m telescope, albeit at low significance.

Observations by the Whipple group 
of Mkn~421 have shown that it is capable of extremely
energetic flares on timescales of hours to days. A flare similar to
that seen on 7 May 1996 by the Whipple telescope (\cite{mce97})
with a flux of
F($>$250GeV) $\sim 5.7 \times 10^{-10}$ photons ${\rm cm}^{-2}{\rm s}^{-1}$
(F($>$2TeV) $\sim 3.8 \times 10^{-11}$ photons ${\rm cm}^{-2}{\rm s}^{-1}$
for a $-$1.56 spectral index and maximum photon energy of 10\,TeV) lasting
two hours would be easily detectable by CANGAROO at a significance of
around $7 \sigma$.

\section{Data sample}

When selecting AGN as observation targets for the 3.8\,m
telescope we have used a number of criteria including the
proximity of the source and measurements from X-ray
and gamma-ray satellite experiments. 
As mentioned earlier,
\cite{ste96}  have suggested that nearby XBLs
are the most promising sources of detectable TeV gamma-ray emission.
Multi-frequency studies of blazar emission (\cite{sam96}) suggest
that XBLs may be more compact and have higher electron energies
and stronger magnetic fields than
their radio selected counterparts. 
The current status of 
ground based TeV observations adds support to this belief.
We present here the results of observations on two XBLs and 
two radio selected BL~Lacs (RBLs).
While RBLs are perhaps less promising as TeV sources, well
placed upper limits in the TeV range could help to confirm fundamental
%source 
differences between XBLs and RBLs. The four sources reported in this
paper are described in the following sub-sections.

\subsection{PKS0521$-$365}
PKS0521$-$365
(RA$_{J2000}$ = $05^{h}23^{m}2.0$, 
Dec$_{J2000}$ =$-36^{\circ}$23'2'', z = 0.055)
is a radio selected BL-Lac. It was first detected as a strong radio
source more than 30 years ago (\cite{bol64}). 
The spectral energy distribution of PKS0521$-$365 shows a double
peaked structure that is typical of blazars (\cite{pia96}). The first
peak in the energy distribution occurs in the far IR ($10^{13}-10^{14}$Hz)
and is assumed to be from synchrotron emission from the electrons in the jet.
The second component, possibly from 
IC emission from the same electron population, 
peaks at around 100MeV.

PKS0521$-$365 was viewed by the EGRET experiment on the CGRO during
1992 from May 14 to June 4 (\cite{lin95}).
 A point source, consistent with the position
of PKS0521$-$365 was detected at a statistical significance of 4.3 $\sigma$, 
with an integral source flux above 100MeV of $(1.8 \pm 0.5) \times 10^{-7}$
photons ${\rm cm}^{-2}{\rm s}^{-1}$. The photon spectrum from the EGRET
observation can be fitted with a single power law \\

$dN/dE = (1.85 \pm 1.14) \times 10^{-8} (E/1 {\rm GeV})^{-2.16 \pm 0.36}$
photons ${\rm cm}^{-2}{\rm s}^{-1}{\rm GeV}^{-1}$\\

{\noindent}The hardness of the EGRET photon spectrum and the proximity of the
source make PKS 0521-365 a candidate source for detectable levels
of TeV gamma-ray emission.

The CANGAROO data set for this object consists of 52 on/off pairs of
observations with a total of 89 hours of on-source and 84 hours of
off-source data.

\subsection{EXO\,0423.4$-$0840}

EXO\,0423.4$-$0840 
(RA$_{J2000}$ = $04^{h}25^{m}50.7$, 
Dec$_{J2000}$ =$-08^{\circ}$33'43'', z = 0.039)
was reported as a serendipitous discovery as part of the high galactic
latitude survey in the 0.05--2.0\,keV range by EXOSAT (\cite{gio91}).
Associating the source with the galaxy, MCG --01--12--005
(incorrectly given in \cite{gio91} as MCG +01--12--005)
and noting the high X-ray luminosity ($>10^{43}$ ergs ${\rm s}^{-1}$) 
of the source, Giommi et al. (1991) proposed the source as a candidate
BL~Lac object.
This would make the source 
the third closest such object known (after Mkn~421 and Mkn~501),
and the closest in the southern hemisphere.
We note however that \cite{kir90} determine a
redshift of 0.0392 for the source they designate IRAS~04235-0840B and
identify with the HEAO source 1H~0422-086, and which they classify as
a type 2 Seyfert.  Clearly higher resolution X-ray studies are
required to firmly establish the identity of this source.
 
We have observed EXO\,0423.4$-$0840 during
October 1996 obtaining a total raw
data set comprising 20 hours of on-source and 17 hours of off-source data.

\subsection{PKS2005$-$489}

PKS2005$-$489 
(RA$_{J2000}$ = $20^{h}09^{m}25.4$, 
Dec$_{J2000}$ = $-48^{\circ}$49'54'', z=0.071)
is an XBL. X-ray measurements of PKS2005$-$489 by EXOSAT show extremely
large flux variations on timescales of hours (\cite{gio90}).
Initially reported as a marginal EGRET detection (\cite{mon95}),
a more accurate background estimation decreased the
significance below the level required for inclusion in the Second
EGRET Catalog (\cite{tom95}).
However the fluxes above 1\,GeV are more significant than those
above 100\,MeV suggesting that the source, 
in a part of the
sky that has received relatively poor EGRET exposure,
may indeed be detectable at higher energies (\cite{lin97}). 

We have observed PKS2005$-$489 during August of 1993 and during 
August/September 
1994 obtaining 41 hours of on-source and 38 hours of off-source data. 

\subsection{PKS2316$-$423}

PKS2316$-$423 
(RA$_{J2000}$ = $23^{h}19^{m}05.8$, 
Dec$_{J2000}$ = $-42^{\circ}$06'48'', z= 0.055) 
is a radio selected BL~Lac object. Assuming a magnetic field
strength of $B \leq 10^{-3}$G, the emission from radio through to X-ray
wavelengths is consistent with synchrotron radiation from electrons with
$E > 10^{13}$eV (\cite{cra94}). PKS2316$-$423 is not detected
by the EGRET telescope on the CGRO. 

The CANGAROO 3.8\,m telescope observed PKS2316$-$423 during July 1996  for
a total of 26 hours of on-source data and 25 hours of off-source data.

\section{Image analysis and Monte Carlo simulations}

Prior to image analysis a data integrity check is performed to
test for the presence of cloud or electronics problems. For a subset
of observations tracking calibration is tested by monitoring changes
in single-fold rates as local stars rotate through the field of view. 
Using this method the pointing direction can be inferred to an
accuracy of around $0.05^{\circ}$.

For each event trigger the tubes associated with the image are
selected using the following criteria:
(i) The TDC of the tube must indicate that the tube has exceeded
its discrimination threshold within 50\,ns around the trigger time of the
event, and
(ii) The ADC signal in the tube must be at least 1 standard
deviation above the RMS
of background noise for that tube.

An image is considered suitable for parameterization if it contains more
than 4 selected tubes, and if the total signal for all tubes in the image 
exceeds 200 ADC counts (around 20 p.e.). About 25\% of raw images
are rejected by these two selection conditions. Surviving images are
parameterized after \cite{hil85}, with the gamma-ray domains for
our observations being :
\\
\\
$0.5^{\circ} <$ Distance $< 1.1^{\circ}$\\
$0.01^{\circ} <$ Width $< 0.08^{\circ}$\\
$0.1^{\circ}<$  Length $< 0.4^{\circ}$\\
$0.4 <$ Concentration $< 0.9$\\
alpha $< 10^{\circ}$\\

The cumulative percentages of events passing each selection condition
are given in table~\ref{cuts}. The 
numbers shown are based on image cuts applied to
Monte Carlo gamma-rays and protons. The efficiency of the cuts for the
Monte Carlo protons is consistent with that seen for real data.

\begin{table}
 \begin{center}
\begin{tabular}{ccc}
\hline
& & \\
Cut  & Gammas &  Protons \\
& & \\ \hline 
Raw data      & 100\% & 100\% \\
ADC sum$>$200 &  90.0 & 77.5 \\
Distance      &  65.4 & 44.6 \\ 
Width         &  54.8 & 20.4 \\
Length        &  54.0 & 16.5 \\
Concentration &  40.3 & 10.9 \\
alpha         &  38.8 &  1.3 \\
\hline  
\end{tabular}
\end{center}
\caption{Cumulative percentages of events passing gamma-ray
selection cuts.\label{cuts}}
\end{table}

To estimate fluxes and upper limits for our data we use an exposure
calculation based on a detailed Monte Carlo simulation of the response
of our telescope to gamma-ray initiated EAS. Simulations have been 
based on the MOCCA simulation package (\cite{hil82}) which models all relevant
particle production processes and includes atmospheric absorption effects
for the \v Cerenkov photons that are produced. The energies of simulated
gamma-ray primaries were selected from a power law with integral exponent
$-$1.4, with a minimum primary energy of 500\,GeV. Core distances were
selected from an appropriate distribution out to a limiting core
distance of 250\,m. Simulated gamma-ray images were then subjected to the
same selection criteria as the real data. Fluxes and upper limits are 
calculated by comparing measured gamma-ray excess rates to those predicted
by the simulations. If we assume that our telescope model is
correct, this method of flux estimation has only two free parameters --- 
the source spectral index and source cutoff energy. 
For total flux calculations the estimate of cutoff energy is not
critical --- the generally steep nature
of gamma-ray source spectral indices ensures that the bulk of the
flux is around the threshold energy. 

\section{Results}

\begin{figure}
 \picplace{65mm}
 \includegraphics{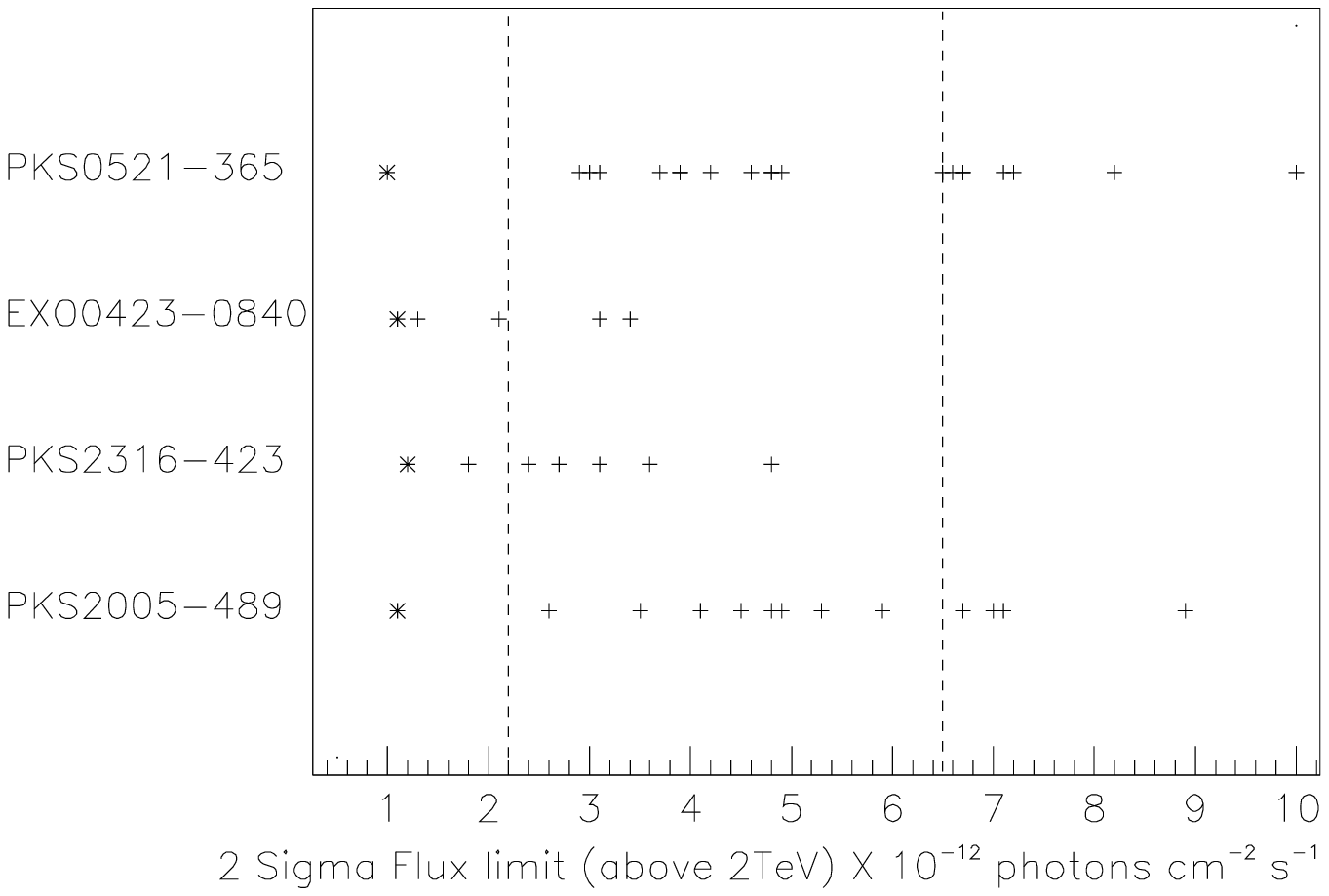}
   \caption{Scatter plot of night by 
    night $2\sigma$ flux limits for each of the
    sources (indicated by crosses). Also shown are the $2\sigma$ flux limits
    for the total data set for each source (stars). The two dashed lines
show the $>$2TeV fluxes from Mkn~421 (left) and Mkn~501 (right) (see text
for details).}
 \label{flux}
\end{figure}

The total data set for each source has been tested for the presence
of gamma-ray signals. 
The significance of gamma-ray excesses have been calculated using
a method based on that of \cite{li83}:

\begin{displaymath}
S  = \sqrt{2} \left\{ N_{on}  
{\rm ln} \left[\frac{1+\beta}{\beta} 
\left( \frac{N_{on}}{N_{on}+N_{off}} \right) \right] \right. 
\end{displaymath}

\begin{equation}
 \ \ \ \ \ \ \ \ \ \ \  \left. + N_{off} {\rm ln} 
\left[ (1+\beta) 
\left( \frac{N_{off}}{N_{on}+N_{off}} \right) \right]  \right\}^{1/2}
\end{equation}

\noindent where S is the statistical significance and
$\beta$ is the ratio of events in the on-source observation to
those in the off-source observation in the range 
$30^{\circ} < {\rm alpha} < 90^{\circ}$ (where alpha is the image parameter
describing image orientation). The values of $N_{on}$ and $N_{off}$
are calculated from the gamma-ray selected (alpha $< 10^{\circ}$)
data for the on and off-source observations respectively. 
This method tends to slightly 
overestimate the significances because it does not
account for the statistical uncertainty
in calculating the value of $\beta$. Based on Monte Carlo simulations
we estimate that this effect is small --- and less than 
the typical systematic
differences caused by the variations in parameter distributions between
on and off-source observations. 
In the total data set for each source no significant excess is seen for
those events in the gamma-ray domain.
The calculated excesses are 
$-0.27 \sigma$ (PKS0521$-$365),
$-0.99 \sigma$ (EXO\,0423.4$-$0840),
$-0.91 \sigma$ (PKS2005$-$489) and
$+0.22 \sigma$ (PKS2316$-$423).
Upper limits to steady emission have
been calculated after \cite{pro84} and are shown in the scatter plot 
in Fig.~\ref{flux}. 

\begin{table}
 \begin{center}
\begin{tabular}{cccc}
\hline
& & & \\
PKS0521 & PKS2005 &  PKS2316 & EXO0423\\
MJD\, \, \,  F & MJD\, \, \,  F & MJD\, \, \, \,  F & MJD\, \, \, \,  F \\ \hline 
9655.7\, \,   2.9 & 9220.5\, \,   4.5 & 10304.8\, \,   4.8 & 10367.7\, \,   2.1   \\
9663.7\, \,   6.5 & 9221.6\, \,   5.3 & 10305.7\, \,   2.7 & 10368.7\, \,   1.3 \\ 
9684.6\, \,   6.7 & 9222.6\, \,   7.1 & 10306.7\, \,   3.1 & 10369.7\, \,   3.1  \\
9685.7\, \,   3.9 & 9576.7\, \,   4.9 & 10307.8\, \,   2.4 & 10370.7\, \,   3.4 \\
9686.7\, \,   3.1 & 9577.7\, \,   4.8 & 10308.7\, \,   1.8 & \\
9687.7\, \,   4.8 & 9597.5\, \,   7.0 & 10311.5\, \,   3.6 & \\
9688.7\, \,   3.7 & 9599.5\, \,   2.6 & & \\
9690.7\, \,   4.6 & 9600.5\, \,   3.5 & & \\
9691.7\, \,   7.1 & 9601.5\, \,   5.9 & & \\
9713.6\, \,   4.9 & 9602.5\, \,   6.7 & & \\
9714.7\, \,   10.0 & 9604.6\, \,   8.9 & & \\
9715.6\, \,   7.2 & 9605.6\, \,   4.1 & & \\
9717.6\, \,   4.8 & & & \\
9718.6\, \,   8.2 & & & \\
9719.6\, \,   3.0 & & & \\
9720.5\, \,   4.2 & & &  \\
9722.6\, \,   6.6 & & & \\
\hline   
\end{tabular}
\end{center}
\caption{The 2$\sigma$ flux upper limits, F($>$2TeV) 
($\times 10^{-12}$ photons cm$^{-2}$ s$^{-1}$)  and approximate 
MJD (-40000) for each on-source observation.\label{obs}}
\end{table}

\begin{figure}
 \picplace{60mm}
 \includegraphics{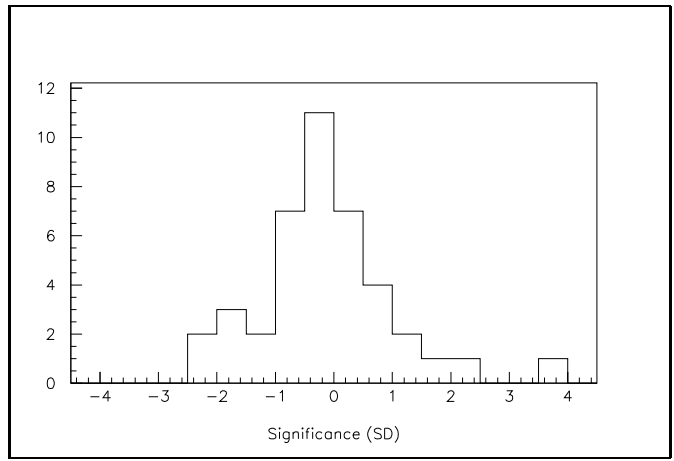}
  \caption{Distribution of the significances of night by night
           excesses for all sources.}
 \label{sig}
\end{figure}

We have also searched our data set for gamma-ray emission on a night by 
night basis. In general our observations of a source consist of a long
(several hours) on-source run, with a similar length off-source
run, offset in RA to provide the same coverage of azimuth and zenith. 
The flare search has been performed by calculating the on-source excess
for each pair of on/off observations each night.
In cases where there
is  no matching off-source run, an equivalent off-source run from another
nearby night is used.
Figure~\ref{sig} shows the distribution of on-source significances
for all three sources. There is no evidence for gamma-ray flares on
the timescale of $\sim$ 1 night for any of the sources.
The most significant nightly excess (from PKS0521-365) has a nominal 
significance of $3.7 \sigma$ but after allowing for the number of 
searches performed this significance is reduced to less than 
$3\sigma$.
The upper limits to
gamma-ray emission for these observations are shown in Fig.~\ref{flux}.
We have also included, in table~\ref{obs}, a list of 
upper limits to gamma-ray emission for each individual observation.

\section{Discussion}
The interpretation of upper limits from BL~Lacs is difficult, because at
the present time there is no detailed model to predict TeV fluxes from
the different classes of BL~Lacs.  
\cite{ste96} have 
attempted to predict TeV fluxes from a number of nearby XBLs by assuming
that all XBLs have similar spectral characteristics to the known TeV
emitter Mkn~421. \cite{mac95} argue that the observed X-ray/TeV gamma-ray
flux increases from Mkn~421 during flares indicate
that the TeV gamma-rays are produced
by a synchrotron self-Compton (SSC) mechanism. 
\cite{ste96} note that the Compton emission from the
electrons in the jet has a similar spectrum to the synchrotron
component, but upshifted by the square of the 
maximum Lorentz factor (estimated to be $\sim 10^{4.5}$ for Mkn~421)
of the electrons. For Mkn~421 and PKS2155$-$304
(both XBLs detected by EGRET) the luminosities in the Compton and
synchrotron regions are nearly equal. Assuming that other XBLs
also have $L_{C}/L_{syn} \sim 1$, \cite{ste96} derive the following
relationship
between X-ray and TeV source fluxes :\\

$\nu_{TeV}F_{TeV} \sim \nu_{x}F_{x}$\\

{\noindent}where $\nu$ is the frequency and $F$ the flux in each energy band.
Using an assumed $E^{-2.2}$ spectral index for all XBLs, and absorption
of TeV gamma-rays in the intergalactic infra-red based on an average of
Models 1 and 2 from \cite{ste97}, they estimate TeV fluxes above 1\,TeV
from a number of nearby XBLs based on EXOSAT X-ray flux measurements.
It should be noted that a more recent paper (\cite{ste98}) concludes that
the absorption of TeV gamma-rays in the infra-red background is 
overestimated in \cite{ste97}. Allowing for this, the predicted flux
for PKS2005$-$489  above 1\,TeV is
F($>1 {\rm TeV})=1.3 \times 10^{-12}$ photons ${\rm cm}^{-2}{\rm s}^{-1}$.
Above 2\,TeV (the threshold of CANGAROO in this analysis) this flux would
be F($>2{\rm TeV})=0.34 \times 10^{-12}$ photons ${\rm cm}^{-2}{\rm s}^{-1}$, 
below the flux sensitivity of the measurement made in this paper. 
The photon spectral index assumed in \cite{ste96}
is now incompatible with recent measurements of Mkn~421 and Mkn~501.
It is also possible that the SSC mechanism is not primarily 
responsible for
the TeV gamma-ray emission and a number of other mechanisms have 
been suggested (\cite{man93,der94} and references therein).

Of the other predictions of \cite{ste96},
\cite{ker95} derive
an upper limit to the $>$0.3~TeV emission
for the XBL 1ES\,1727+502 that is above the calculated flux.
It is also worth noting that a deep exposure of the RBL BL~Lacertae
with the Whipple telescope did not yield a detection
(\cite{cat97b}).

Current and future observations of a range of
BL~Lacs by ground based \v Cerenkov telescopes should help to clarify
the mechanisms for the production of high energy photons in these sources.

\section{Conclusion}
Analysis of CANGAROO data shows no evidence for long-term or short-term
emission
of gamma-rays above 2\,TeV from the 
BL~Lacs PKS0521$-$365, EXO\,0423.4$-$0840,
 PKS2005$-$489, PKS2316$-$423. The $2\sigma$ upper
limits to steady emission are 
$1.0 \times 10^{-12} {\rm cm}^{-2}{\rm s}^{-1}$ (PKS0521$-$365),
$1.1 \times 10^{-12} {\rm cm}^{-2}{\rm s}^{-1}$ (EXO\,0423.4$-$0840),
$1.1 \times 10^{-12} {\rm cm}^{-2}{\rm s}^{-1}$ (PKS2005$-$489), and
$1.2 \times 10^{-12} {\rm cm}^{-2}{\rm s}^{-1}$ (PKS2316$-$423).
For the XBL PKS2005$-$489 the flux limits presented in this paper
 do not constrain the TeV emission levels predicted by the simple model
of \cite{ste96}. 

\begin{acknowledgements}
The authors would like to thank O. de Jager for helpful comments.
This work is supported by a Grant-in-Aid in Scientific Research from the
Japan Ministry of Education, Science and Culture, and also by the 
Australian Research Council and the International Science and 
Technology Program. MDR acknowledges the receipt of a JSPS fellowship from
the Japan Ministry of Education, Science, Sport and Culture. 
\end{acknowledgements}

\end{document}